\documentclass[prb,twocolumn,groupedaddress,showpacs]{revtex4}
\usepackage{graphicx}
\begin{document}
\title{Characterizing the network topology of the energy landscapes of 
atomic clusters}
\author{Jonathan P.~K.~Doye}
\email{jpkd1@cam.ac.uk}
\affiliation{University Chemical Laboratory, Lensfield Road, Cambridge CB2 1EW, United Kingdom}
\author{Claire P.~Massen}
\affiliation{University Chemical Laboratory, Lensfield Road, Cambridge CB2 1EW, United Kingdom}
\date{\today}
\begin{abstract}
By dividing potential energy landscapes into basins of attractions surrounding 
minima and linking those basins that 
are connected by transition state valleys,
a network description of energy landscapes naturally arises.
These networks are characterized in detail for a series of small 
Lennard-Jones clusters
and show behaviour characteristic of small-world and scale-free networks.
However, unlike many such networks, this topology cannot reflect the rules
governing the dynamics of network growth, because they are static spatial 
networks. Instead, the heterogeneity in the networks stems from 
differences in the potential energy of the minima, and hence the hyperareas
of their associated basins of attraction. The low-energy minima with
large basins of attraction act as hubs in the network.
Comparisons to randomized networks with the same degree distribution reveals
structuring in the networks that reflects their spatial embedding.
\end{abstract}
\maketitle

\section{Introduction}

In recent years, characterizing the energy landscape of a system in
order to gain a better understanding of its behaviour
has become an increasingly popular research approach,\cite{Wales03} 
with many applications in the fields of protein folding,\cite{Bryngelson95} 
clusters and supercooled liquids.\cite{StillW84a,Still95,Debenedetti01} 
For example, a common
explanation for how proteins are able to overcome the Levinthal 
paradox\cite{Levinthal_Mark} and
fold to their native state on biologically reasonable time scales
is in terms of a funnel-like energy landscape. 

Here, we want to take a somewhat different approach and focus not
on the relationship between the landscape and a system's behaviour, but on
some of the fundamental properties of potential energy landscapes. 
For example, the number of minima increases exponentially with system size
\cite{StillW82,Still99,Tsai93a}
and the size scaling for higher-order saddle points has also been
obtained.\cite{Doye02a,Shell03,WalesD03}
It is also known that the distribution of minima should be 
a Gaussian function of the potential energy.\cite{Sciortino99a,Buchner99a}

In this paper we seek to provide fundamental new insights
into the structural organization, and particularly the connectivity, 
of an energy landscape. 
To achieve this we first map the landscape onto a network, 
and then analyse the topology of this network for a series of small clusters
for which complete networks can be obtained. 
A brief report of some of this work has already appeared.\cite{Doye02c}

Our analysis of these networks is heavily influenced by the 
literature on complex 
networks.\cite{Strogatz01,Albert02,Linked,Bornholdt03,Dorogovtsev03,Newman03a}
The recent interest in this area was in part sparked by the seminal paper of 
Watts and Strogatz,\cite{Watts98} who showed 
that many real-world networks have features 
that are typical both of a lattice (the presence of local order) and of a 
random graph\cite{Erdos59,Erdos60} (short average separations 
between the nodes), and introduced a `small-world' network model that could 
represent both these features. Subsequently, it has been shown that many 
networks also have a probability distribution for the number of connections
to a node (in network parlance, the degree distribution) that has a power-law 
tail.  Such `scale-free' networks\cite{Barabasi99} have 
been found in an impressively diverse range of fields, 
including astrophysics,\cite{Hughes03} geophysics,\cite{Baiesi04} 
information technology,\cite{Albert99} biochemistry,\cite{Jeong00,Jeong01}
ecology\cite{Dunne02} and sociology.\cite{Liljeros01}

The paper is organized as follows. In Section \ref{sect:eland_net} 
we explain how energy landscapes can be described in terms of
networks. In Section \ref{sect:results} we present a detailed characterization
of these networks. This analysis initially focusses on whether the networks 
show small-world and scale-free behaviour, before going on to 
look at more subtle features of these networks. 
Then, in Section \ref{sect:discuss} we discuss the implications of our results
for understanding the dynamics on complex energy landscapes.

\section{Energy landscapes as networks}
\label{sect:eland_net}

\begin{figure}
\includegraphics[width=8.6cm]{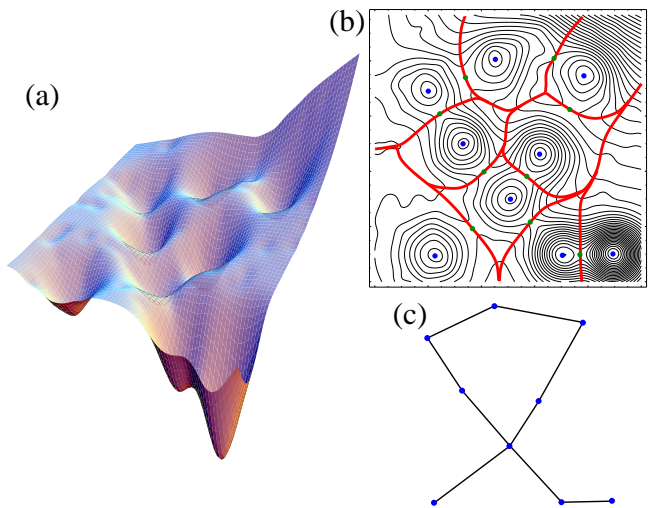}
\caption{(Colour online).
(a) A model two-dimensional energy surface. 
(b) A contour plot of this surface
illustrating the inherent structure partition of the configuration space 
into basins of attraction surrounding minima.  
The basin boundaries are represented by thick lines, 
and the minima and transition states by dots.
(c) The resulting representation of the landscape as a network.
}
\label{fig:PES}
\end{figure}

To model an energy landscape as a network one must first decide on a definition 
both of a state of the system and how two states are connected. 
The states and their connections will then provide the nodes and edges of 
the network. 
For systems with continuous degrees of freedom, 
perhaps the most natural way to achieve this is through the 
`inherent structure' mapping of Stillinger and 
Weber.\cite{StillW82,StillW84a} In this mapping each point in 
configuration space is associated with the minimum (or `inherent structure')
reached by following a steepest-descent path from that point. This mapping
divides configuration space into basins of attraction surrounding 
each minimum on the energy landscape, and is illustrated for a model 
two-dimensional energy landscape in Fig.\ \ref{fig:PES}.

For systems with large numbers of degrees of
freedom, the energy landscape is an extremely complicated 
multi-dimensional function.
The inherent structure approach provides a way of dealing with this
complexity by breaking the landscape up into more manageable pieces. 
For example, good approximations to the partition function of a basin of 
attraction and to the rates of transitions between two basins 
can be obtained within the harmonic approximation 
(and if necessary with additional 
anharmonic corrections), thus allowing 
thermodynamic and dynamic properties of a system to be 
calculated.\cite{Wales03} 

The inherent structure approach also provides a natural partition for the
dynamics of a system. At sufficiently low temperature, the system will spend 
most of its time oscillating within a basin of attraction surrounding a 
minimum with occasional hops between basins along a transition state valley.\cite{Tx}
This interbasin dynamics can then be represented as a walk on a network, 
whose nodes correspond to the minima and where edges link 
those minima that are directly connected by a transition state 
(a first-order saddle point). Such a network is illustrated for the model
surface in Fig.\ \ref{fig:PES},
and we will term them inherent structure networks.

In chemistry such networks are often called reaction graphs, and 
have been used to analyse isomerization in small molecules and 
clusters. In these applications, each permutational isomer of 
a structure is usually considered separately. For example, in 
Ref.\ \onlinecite{Taketsugu02} the reaction graph connecting the eight
permutational isomers of water dimer is depicted. Such knowledge is 
for example important for understanding the pattern of 
splittings that results from tunnelling between these isomers.

For an atomic cluster with a single atom type 
the number of permutational isomers of a minimum is $2N!/h$ where $h$
is the order of the point group of that minimum. 
This factor makes consideration of
the network with a node for each permutational isomer unfeasible 
for all but the smallest sizes.
For example, for a 7-atom Lennard-Jones (LJ) cluster the four geometrically
distinct minima give rise to 8904 permutational isomers, but for LJ$_{14}$
the 4196 geometrically distinct minima give rise to $6.68 \times 10^{14}$
permutational isomers. Hence, here we consider the network where each 
geometrically-distinct minimum is represented by a single node. 
Indeed, this approach is perfectly reasonable, because one is usually 
only interested in the structure of the system, and not in
the permutational order.
The network we consider here will also be unweighted, 
i.e.\ we are only interested in whether two minima are connected, and 
not in the weight, however defined, of this connection. 

\begin{table*}
\caption{\label{table:LJ}
Properties of the energy landscapes and 
inherent structure networks for LJ$_N$ for $N=7-14$.
$n_{\rm min}$ is the number of minima;
$n_{\rm ts}$ is the number of transition states; 
$n_{\rm edges}$ is number of edges in the network; 
$p$ is the average probability of an edge between any two nodes;
$\langle k \rangle$ is the average degree; 
$k_{\rm max}$ is the maximum degree;
$l_{\rm ave}$ is the average separation between nodes; 
$r$ is the radius of the network;
$d$ is the diameter of the network;
$C_1$ and $C_2$ are different definitions of the clustering coefficient;
and $r_k$ and $r_E$ are the assortativity 
coefficients\cite{Newman02a,Newman03b} 
with respect to the degree and the potential energy of the minimum, 
respectively.
The uncertainties in the last two quantities have been calculated using the 
jack-knife method.}
\begin{ruledtabular}
\begin{tabular}{cccccccccccccc}
$N$ & $n_{\rm min}$ & $n_{\rm ts}$ & $n_{\rm edges}$ & $p$ & $\langle k \rangle$ & $k_{\rm max}$ & 
 $l_{\rm ave}$ & $r$ & $d$ & $C_1$ & $C_2$ & $r_k$ & $r_E$ \\
\hline
 7 &    4 &    12 &      5 & 0.833 &  2.50 &    3 & 1.167 & 1 & 2 & 0.750 & 0.833 & $-0.667\pm0.359$ & $-0.559\pm0.321$ \\
 8 &    8 &    42 &     16 & 0.571 &  4.00 &    6 & 1.464 & 2 & 3 & 0.621 & 0.604 & $-0.441\pm0.236$ & $-0.379\pm0.250$ \\
 9 &   21 &   165 &     74 & 0.352 &  7.05 &   15 & 1.714 & 2 & 3 & 0.504 & 0.607 & $-0.287\pm0.130$ & $-0.114\pm0.114$ \\ 
10 &   64 &   635 &    359 & 0.178 & 11.22 &   34 & 2.146 & 3 & 4 & 0.421 & 0.519 & $-0.018\pm0.058$ &  $0.080\pm0.054$ \\
11 &  170 &  2424 &   1623 & 0.113 & 19.09 &   86 & 2.135 & 3 & 4 & 0.292 & 0.362 & $-0.081\pm0.024$ &  $0.101\pm0.026$ \\
12 &  515 &  8607 &   5854 & 0.044 & 22.73 &  281 & 2.300 & 3 & 5 & 0.183 & 0.306 & $-0.100\pm0.009$ &  $0.097\pm0.013$ \\
13 & 1509 & 28756 &  20708 & 0.018 & 27.45 &  794 & 2.394 & 3 & 5 & 0.110 & 0.260 & $-0.097\pm0.003$ &  $0.072\pm0.007$ \\
14 & 4196 & 87219 &  61085 & 0.007 & 29.12 & 3201 & 2.325 & 3 & 6 & 0.052 & 0.249 & $-0.066\pm0.001$ &  $0.082\pm0.004$ \\ 
\end{tabular}
\end{ruledtabular}
\end{table*}

In the first instance, self-connections and multiple edges are also excluded, 
even though there are transition states that mediate degenerate 
rearrangements between permutational isomers of the same minimum, and
there can be more than one geometrically distinct transition state 
connecting two particular minima. 
As a consequence of these exclusions the number of edges in our networks, 
so defined, is roughly 30\% less than the total number of transition states
for the larger clusters (See Table \ref{table:LJ}). 
We exclude self-connections, both because 
we are not interested in permutational order, and because degenerate 
rearrangements make no contributions to the structural dynamics, 
and multiple edges because we are not so much interested in the number of 
connections between two minima, but just whether they are connected.
Besides, if we were concerned with the number of connections it would not be
as simple as counting the number of geometrically distinct transition states
that connect the two minima because of symmetry factors. 
The number of versions of a transition state
that connects to a minimum for a non-degenerate rearrangement is given by 
the factor $h_{\rm min}/h_{\rm ts}$, 
i.e.\ the ratio of the orders of the point groups of the minimum and the
transition state.\cite{Wales03} 

The properties of these types of networks have effectively been 
studied before, albeit in a directed form where the links have 
been weighted by the rate of transition from one minimum to another 
at a particular temperature.\cite{BerryK95,Miller99b,Doye99a,Wales02}
This approach leads to a rate matrix from 
which an exact solution of the interminimum dynamics of the
system can be obtained in terms of the eigenvalues and eigenvectors of 
this matrix using a master equation method.
However, in all these applications the emphasis was on the 
resulting dynamics and not on the topological structure of the networks. 
Moreover,
because the focus was the dynamics of a particular transition and not the 
global dynamics, most of the networks studied were incomplete. 

We should mention three other studies that have sought to characterize 
the connectivity of configuration space in terms of complex 
networks.\cite{Amaral00,Scala01b,Wuchty03}
The first was of a two-dimensional lattice polymer, where each discrete 
configuration was a node in the network and configurations were linked to those
that could be obtained by the application of a single 
elementary move using a move set of corner flips
and ``crankshaft'' rotations.\cite{Amaral00,Scala01b}
The second was of RNA considered at the secondary structure 
level, where a node corresponded to those configurations that had
the same map of contacts between nucleotides and nodes were linked 
to those that could be obtained by an elementary change in the contact 
map.\cite{Wuchty03}
The main differences between these networks and our own is in the discrete,
rather than continuous, nature of the configuration space and the definition 
of a state---there is no equivalent of the inherent structure mapping 
that groups configurations into a single state (or basin). We will
discuss the consequences of these differences later in the paper, 
when we report the properties of the inherent structure networks.
The third study,
which examined the connectivity of the conformation space of 
polypeptides,\cite{Rao04} 
is in a much more similar spirit to the current work, because the 
space is continuous and connections have a dynamical basis.
Given the much larger size of the configuration space than in the current
study, a more coarse-grained picture of the landscape is required. Rao and
Caflisch probe the connections between {\it free energy} minima defined
through structural order parameters that assign a secondary structure
character to each amino acid. The sampling of the landscape is not
complete, but instead based on long molecular dynamics simulations at
a temperature near to the folding transition. The properties of these
networks are much more similar to the present networks.

\section{Network Properties}
\label{sect:results}

We chose to characterize the inherent structure networks of a series of 
Lennard-Jones clusters, whose potential energy is given by 
\begin{equation}
V = 4\epsilon \sum_{i<j}\left[ \left(\sigma\over r_{ij}\right)^{12} - \left
(\sigma\over r_{ij}\right)^{6}\right],
\end{equation}
where $\epsilon$ is the pair well depth and $2^{1/6}\sigma$ is the
equilibrium pair separation. We concentrate on small systems for which 
we are able to characterize the energy landscape completely, because
we are interested in the global topology of the complete inherent 
structure network. We choose to study clusters, rather than a bulk system
to avoid complications associated with boundary conditions.
For clusters we do not have to worry about the density dependence of the
networks. Moreover, for systems that are small enough for the complete 
inherent structure network to be sampled, periodic boundary conditions
represent a significant perturbing constraint on the system, and so do not
provide an accurate representation of a bulk system. By contrast, small
LJ clusters are interesting systems in their own right, whose structural,
thermodynamic and dynamic properties have been well characterized.

A (near-)complete sampling of the relevant stationary points of 
LJ$_{N}$ up to $N$=13
had previously been obtained in a work looking at the scaling behaviour of the
numbers of stationary points.\cite{Doye02a}
In addition we generated samples of minima and transition 
states for LJ$_{14}$. 

Here, we quickly summarize the approach used.
It is an iterative procedure whereby starting from each minima in
our sample we perform a series of transition state searches. We then step off
any new transition states along the direction of negative curvature
to find the connecting minima. More transition state searches are then 
performed from any new minima. This procedure is repeated until the number of
transition states appears to have converged with respect to increasing numbers
of searches. However, to find all the 
transition states we also have to perform searches for second-order 
saddle points starting from the transition states and minima. 
From these second-order saddle points we then repeatedly perform 
transition state searches. With this additional tool the number of 
transition states increases significantly and does appear to converge to
the true total, although of course there is no proof of this.

The number of minima and transition states found by this procedure 
for each cluster size are given in Table \ref{table:LJ}.
The factor that limits the size for which such complete stationary point 
samples can be generated is not so much the number of stationary points,
but the huge number of searches that need to be performed to achieve 
convergence. We were able to map out the complete inherent structure
networks for all clusters up to $N$=14. The network data is available 
online.\cite{ISN_online}

There is a theoretical expectation that the number of minima and transition
states scale with size as \cite{StillW82,Still99,Doye02a} 
\begin{equation}
n_{\rm min}=e^{\alpha N}\qquad\hbox{and}\qquad
n_{\rm ts}=aNe^{\alpha N}
\end{equation}
where $\alpha$ and $a$ are constants. These forms are based on the 
assumption of a bulk system (i.e.\ large $N$ and no surface) that
can be divided up into independent equivalent subsystems.
Hence, one should not expect these forms to hold too precisely for small
clusters. Expressions for the number of higher-order saddle points
can also be derived.\cite{Shell03,WalesD03} 
Based on the Hessian index for which the number of stationary points
is a maximum,
a value of $a$=2 appears appropriate for clusters.\cite{WalesD03} 

In network parlance, the number of connections to a node is called the 
degree and denoted by $k$.
Taking the above forms for $n_{\rm min}$ and 
$n_{\rm ts}$ as appropriate and ignoring for the moment that not
all transition states give rise to edges in the networks (because of our
exclusion of multiple edges and self-connections), one expects that
the average degree for our networks should follow
\begin{equation}
\label{eq:avek}
\langle k\rangle\approx{2 n_{\rm ts}\over n_{\rm min}}=2aN
={2a \over \alpha}\log n_{\rm min},
\end{equation}
i.e.\ $\langle k\rangle$ should scale linearly with the size of the clusters, 
or logarithmically with the size of the network. It follows that $p$, 
the fraction of all possible edges in the network that are 
actually connected is given by 
\begin{equation}
p\approx{2 n_{\rm ts}\over n_{\rm min}(n_{\rm min}-1)}
 ={2aN\over e^{\alpha N}-1}
 ={2a \log n_{min}\over \alpha (n_{\rm min} -1)}.
\end{equation}
Hence the networks should become more sparse as they increase in size.

In line with these theoretical expectations, 
significant increases of $\langle k\rangle$ with $N$ and decreases of $p$ with
$N$ are apparent from Table \ref{table:LJ}.
As we will be comparing networks generated from clusters of different size,
it should be remembered that these basic features of the networks are changing. 

\subsection{Small world properties} 
\begin{figure}
\includegraphics[width=8.6cm]{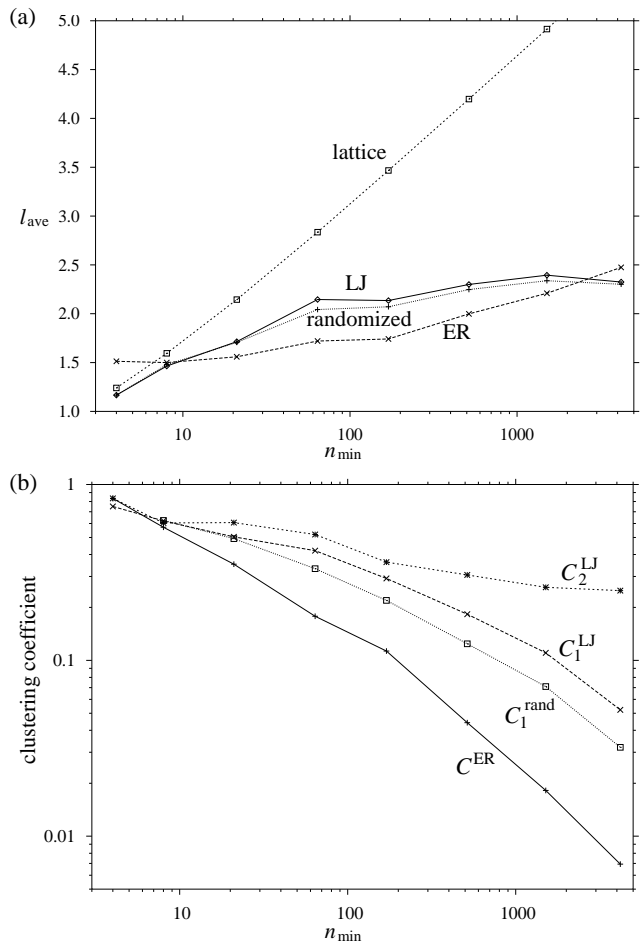}
\caption{
The dependence of (a) the average separation between nodes (in steps) and 
(b) the clustering coefficient, $C$, 
on the size of the network, $n_{\rm min}$. 
The data points for LJ clusters with from 7 to 14 atoms
are compared to the values expected for a lattice (Eq.\ (\ref{eq:lave_latt})) 
and an Erd\H{o}s-Renyi (ER) random graph 
(Eq.\ (\ref{eq:lave_rand})), and to those for
randomized (rand) networks with the same degree distribution as the clusters, 
as labelled. Values of both $C_1$ and $C_2$ are plotted for the LJ clusters.
}
\label{fig:SW}
\end{figure}

One of the key properties when deciding whether a network behaves as a 
`small world' is the average separation between nodes, $l_{\rm ave}$.
More precisely, for each pair of nodes in the network
the shortest pathway between them is found and then the average 
of the number of steps in these pathways is taken. 
Values of $l_{\rm ave}$ are given in Table \ref{table:LJ} and its
dependence on network size is depicted in Fig.\ \ref{fig:SW}(a).
$l_{\rm ave}$ seems to be growing sub-logarithmically with $n_{\rm min}$, 
although the growth is non-monotonic---such size effects are typical of 
small clusters.\cite{Jortner}

Two other quantities that reflect the properties of the shortest 
pathways between nodes are given in Table \ref{table:LJ}. 
The diameter, $d$, is the number of steps in the longest such pathway, and 
the radius, $r$, is 
the number of steps in the longest pathway for the node for 
which this is a minimum, 
i.e.\ all the other nodes are within $r$ steps
of that node. The small values of $l_{\rm ave}$, $r$ and $d$ would suggest
that the network is showing the `small world' effect, but a more sophisticated 
analysis is also needed. 

The usual approach to decide whether the average separation between nodes 
shows behaviour typical of a random graph or of a lattice is to test
whether $l_{\rm ave}$ is a logarithmic 
($\log n_v$) or power-law ($n_v^{1/d}$ where
$d$ is the dimension of the lattice) function of the network size, $n_v$, 
respectively. However, the situation is not that simple for our networks, 
because both the dimension of configuration space 
and the average degree increases with the network size.
Therefore, to test for lattice-like behaviour we need to compare
the network, not to a lattice of fixed dimension and increasing extent, 
but of fixed extent and increasing dimension. It is of fixed extent because
the displacement of an atom that is required to reach the nearest permutational
isomer of any minimum should always be of the order of the atomic size.

For a $3N$-dimensional hypercubic lattice with $L$ lattice points along each 
coordinate direction, the average separation between lattice points 
is approximately the sum of the average difference for each of 
the $3N$ coordinates. More specifically,
\begin{equation}
l_{\rm ave}^{\rm latt}={n_{\rm min}\over n_{\rm min}-1} 3N\, l_{\rm ave}^{1D}, 
                      \approx 3N\, l_{\rm ave}^{1D}, 
\end{equation}
where the initial factor is to exclude the zero length paths between 
a node and itself from the average and
\begin{equation}
l_{\rm ave}^{1D}={1\over L^2} \sum_{i,j=1}^L |i-j|={(L-1)(L+1)\over 3L}.
\label{eq:l1D}
\end{equation}
Hence $l_{\rm ave}^{\rm latt}=N(L-1)(L+1)/L$.\cite{PRLbug}
Therefore, the average separation between lattice points scales linearly with
the dimension of the system. The dependence
of $l_{\rm ave}^{\rm latt}$ on the number of lattice points can be obtained
by substituting for $N$ in the above equation using $n_{\rm latt}=L^{3N}$,
leading to 
\begin{equation}
l_{\rm ave}^{\rm latt}={(L-1)(L+1) \over 3L \log L} \log n_{\rm latt} .
\label{eq:lave_latt}
\end{equation}
Thus, even in the case of a lattice, the average separation scales 
logarithmically, because of the increasingly high dimensionality of 
configuration space.

\begin{figure}
\includegraphics[width=8.6cm]{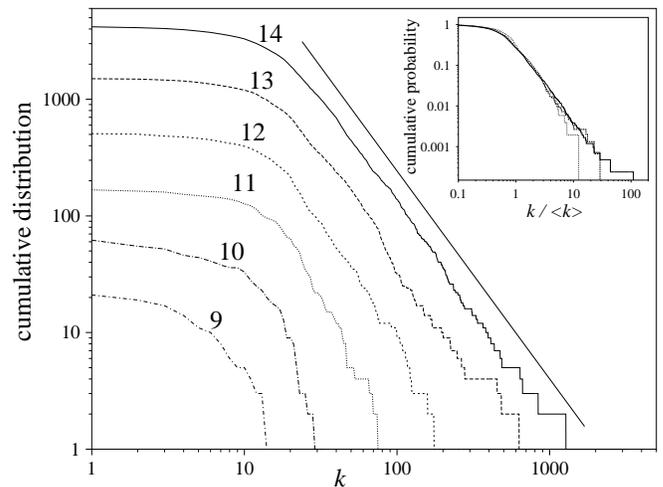}
\caption{
The cumulative distribution for the number of nodes that have more
than $k$ connections. The curves correspond to clusters of different sizes, as labelled.
An additional straight line with slope $-(\gamma-1)$, where $\gamma$=2.78, has
been plotted to emphasise the power law tail.
In the inset the cumulative probability distribution for the 12-, 13- and 14-atom clusters
is plotted against $k/\langle k \rangle$, to 
bring out the universal form of the distribution.
}
\label{fig:degree}
\end{figure}

For a random graph 
\begin{equation}
l_{\rm ave}^{\rm ER}={\log n_v\over\log\langle k\rangle}, 
\label{eq:lave_rand}
\end{equation}
so for our
networks, if this equation were obeyed we would expect a sublogarithmic
scaling with size. In fact, if we substitute the result from Eq.\ 
(\ref{eq:avek}) we get 
\begin{equation}
l_{\rm ave}={\log n_{\rm min}\over\log\left(2a/\alpha\right)+
             \log\left(\log n_{\rm min}\right) }
\label{eq:lave_land}
\end{equation}
The apparent sublogarithmic behaviour of Fig.\ \ref{fig:SW}(a) would 
suggest that as with the Watts-Stogatz small-world networks, the scaling of
$l_{\rm ave}$ is similar to a random graph, but more conclusive evidence 
in favour of this conclusion comes if we compare 
Equations (\ref{eq:lave_latt}) and (\ref{eq:lave_rand}) to our data. 
To apply Eq.\ (\ref{eq:lave_latt}) a value for $L$ was first obtained using
the number of minima for LJ$_{14}$, which gives $L=1.220$.
A much better fit to our data is obtained from the random-graph expression,
confirming our initial assessment.

This scaling behaviour may seem somewhat surprising, since the connections 
between minima are based on the adjacency of the associated basins 
in configuration space, which would perhaps initially suggest a more 
lattice-like picture of configuration space. 
Furthermore, there are no obvious equivalents of the 
random linkages between distant nodes 
that cause the small-world behaviour in the 
Watts-Strogatz networks.\cite{Watts98}
As we will see later, the origin lies elsewhere.

Another feature of many networks is a strong degree of local structure,
as measured by the clustering coefficient. 
The clustering coefficient of node $i$, $c_i$, is defined as the probability 
that a pair of neighbours of $i$ are themselves connected.
However, when extending this concept to the whole graph there are two 
definitions in common usage. The first, $C_1$, is simply the probability that 
any pair of nodes that have a common neighbour are themselves 
connected,\cite{Watts98} and the second, $C_2$, is the average of the local 
clustering coefficient: $C_2=\sum_i c_i/N_{\rm min}$. 
The difference between these two definitions is the relative weight
given to nodes with different degree. High degree nodes make a larger
contribution to $C_1$ because there will be more pairs of nodes that
have a high-degree node as a common neighbour, whereas all
nodes contribute equally to $C_2$. Typically, $C_1<C_2$
because, as is the case here, higher degree nodes tend to have
lower values of $c_i$.

Values of $C_1$ and $C_2$ are given in Table \ref{table:LJ}
and are compared to that for an Erd\H{o}s-Renyi random graph 
($C^{\rm ER}=p$) in Figure \ref{fig:SW}(b). 
It is apparent that the networks for the larger clusters have 
significantly more local structure than the random graphs. 
For LJ$_{14}$
$C_1$ and $C_2$ are 7.53 and 35.90 times larger than $C^{\rm ER}$, respectively.
At small cluster sizes $C\approx C_{\rm ER}$ mainly due to the large values
of $p$---there is going to be little difference from a random graph when the 
network is almost fully connected---but as the networks become more sparse, 
they show increasing local correlations. This feature of the networks is
unsurprising given that the connections are based on adjacency in
configuration space.

\subsection{The degree distribution} 
\label{subsect:degree}
Another very important property of a network is its degree distribution.
These distributions are illustrated for our networks in Figure \ref{fig:degree}.
Remarkably, as the size of the clusters increase a clear power-law tail 
develops. Such networks are typically said to be scale-free, 
even though the power-law
behaviour does not extend back to small values, and so the distribution is 
not totally invariant to rescaling.\cite{Tsallisfit} 
In fact, the initial flat portion of the 
cumulative degree distributions in Fig.\ \ref{fig:degree} implies that
the degree distribution itself has a maximum that lies near to, but below,
$\langle k \rangle$.
Interestingly, the inset to Fig. \ref{fig:degree} suggests that 
the distribution is tending to a universal form independent of cluster size.
The value of the exponent in the power law is 2.78, which is quite 
similar to other scale-free networks (most are between 2 and 3).

\begin{figure}
\includegraphics[width=8.6cm]{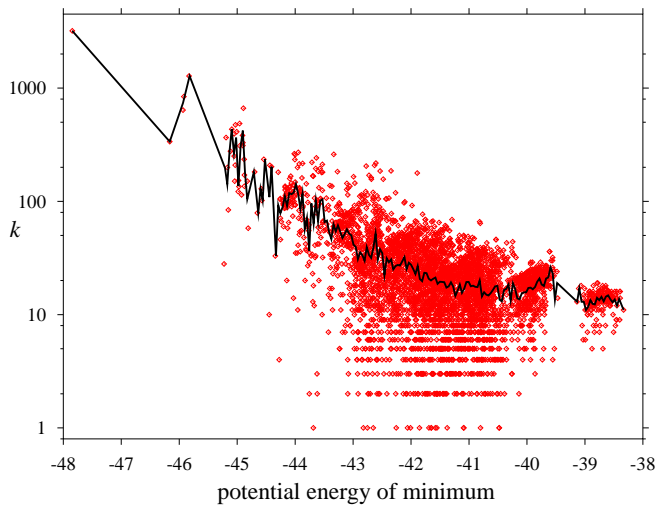}
\caption{(Colour online).
The dependence of the degree of a node on the potential energy 
of the corresponding minimum for 
LJ$_{14}$.
The data points are 
for each individual minimum and the solid line is a binned average.
}
\label{fig:LJ14_kE}
\end{figure}

This result implies that the networks are very heterogeneous, 
with a few hubs having a very large number of connections, but with most
having something near to the average. It also explains the origin of the 
small-world behaviour, as scale-free networks are known to have very
small values of $l_{\rm ave}$. 
This is because the hubs provide natural conduits for short paths between nodes.
In fact, the scaling of $l_{\rm ave}$ with network size 
can be even slower than for Erd\H{o}s-Renyi random 
graphs.\cite{Cohen02,Chung02}

The obvious next question to ask is what features differentiate those
minima with large numbers of connections from those with only a few.
What is the source of the heterogeneity?
Most of the models devised to explain the scale-free properties of
networks are dynamic in nature. For example, in Barabasi and Albert's model
it arises from the preferential attachment of new nodes to those with higher
degree at each step in the growth process.\cite{Barabasi99} 
However, our inherent structure networks are static in nature, and are just 
determined by the form of the interatomic potential and the number of atoms. 

One of the few scale-free models not involving network growth is the 
fitness model of Calderelli {\it et al.},
in which each node is assigned a fitness parameter that controls its 
properties.\cite{Caldarelli02}
For a suitable combination of the distribution of the fitness parameter and
the fitness-dependent linking probability, scale-free networks can result.

The most obvious parameter that might control the properties of our nodes
is the potential energies of the associated minima. 
In Figure \ref{fig:LJ14_kE} we show the dependence of the degree of a node
on its potential energy for our largest network. 
There is a strong correlation; the global minimum is the most connected and
the degree decays to lower values as the energy increases.
This picture is typical of the data for other clusters, and 
only for LJ$_{9}$ and LJ$_{10}$ is the global minimum not 
the most highly-connected hub; in fact it is the fifth lowest-energy minimum
in both these exceptions.
The $k_{\rm max}$ values in Table \ref{table:LJ} indicate the highly-connected
nature of the hubs. The biggest hub is always connected to more than
50\% of the network, and the particularly high value for LJ$_{14}$ (76\%)
explains the non-monotonic behaviour of $l_{\rm ave}$ at $N=14$.

\begin{figure}
\includegraphics[width=8.6cm]{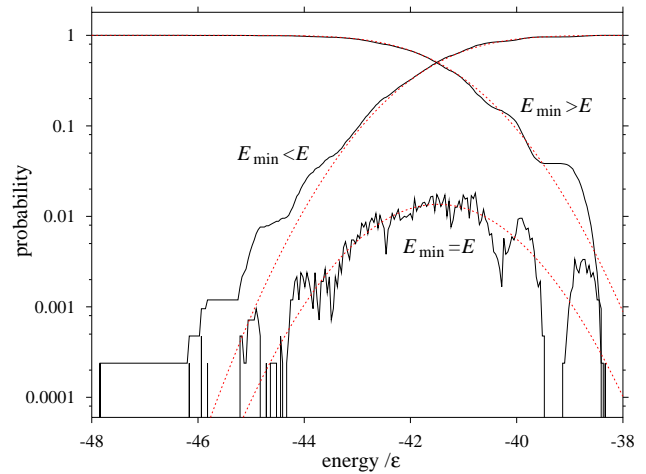}
\caption{(Colour online). 
(Cumulative) probability distributions for the energy of the 
minima for LJ$_{14}$. The distributions are compared to Gaussian 
forms (dotted lines).}
\label{fig:pE}
\end{figure}

The combination of $k(E)$ illustrated in Fig.\ \ref{fig:LJ14_kE} 
with the probability distribution for the potential energy of the minima, 
$p(E)$, must lead to a power-law tail for $p(k)$, and so this might
be one avenue to explain the scale-free behaviour of our networks. 
Indeed, the form of $p(E)$ has been extensively investigated for a
variety of systems, and a physical motivation given for its form.
Based on the minima sampled in the liquid state of bulk matter,
the distribution of minima has been found to be a Gaussian function of
the potential energy.\cite{Sciortino99a,Buchner99a}
This form can be justified from the central limit theorem if it is
assumed the system can be divided into independent subsystems.\cite{Heuer00}

Our data can also be well-approximated by a Gaussian (Figure \ref{fig:pE}),
however there are deviations in the tail regions. In particular, $p(E)$ decays
more slowly than a Gaussian at low energy. This deviation is not so 
surprising, since the analysis is expected to be most applicable where 
there is a quasicontinuous distribution of minima, rather than for the
more discrete spectrum of minima in the low-energy 
tail that is associated with the solid form of the clusters. However,
this part of the distribution is particularly significant as it determines the
properties of the highly-connected minima in the scale-free tail. Indeed,
a more detailed analysis shows that the non-Gaussian nature of the low-energy
tail of $p(E)$ combines with the faster than exponential increase of $k(E)$
to produce the power-law tail for $p(k)$.
However, it is no more clear why these functions should have these precise
forms than why $p(k)$ has a power-law tail, so relatively little new physical
insight has been obtained.

Probably, a more fruitful way of understanding the heterogeneous nature
of the networks is by thinking in terms of the hyperareas of the 
basins of attraction surrounding a minimum. 
The high connectivity of the low-energy minima is likely 
to stem from their large basins of attraction, because this gives them long
basin boundaries, and hence the likelihood of a greater number of 
transition states on these boundaries. Indeed, there is evidence that for 
LJ$_{38}$ the basins of attraction rapidly decrease in area with 
increasing potential energy.\cite{Doye98e} Some statistics are also available 
for the probability of reaching the global minima of LJ clusters from
a random starting point, which provide a direct measure of the relative area
of the global minimum.\cite{StillD90b,Bytheway} These studies indicate that 
this probability is significantly larger than $1/n_{\rm min}$, and
thus that they have large basins of attraction.
We intend to pursue this idea further in future work, in particular to examine
in detail whether the relationship between basin area and connectivity is as
we have suggested.

\begin{figure}
\includegraphics[width=8.6cm]{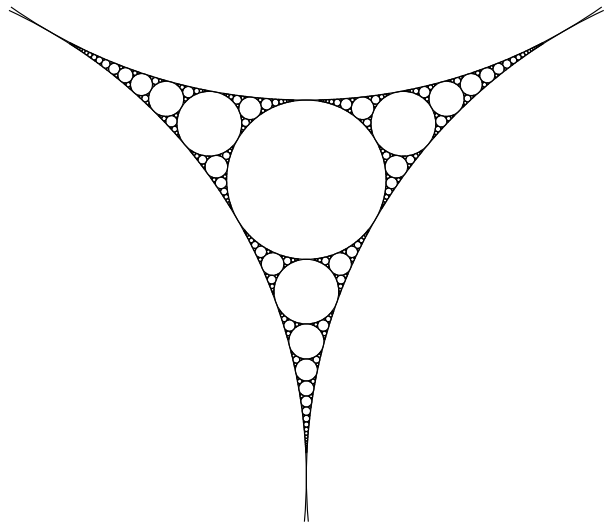}
\caption{An Apollonian packing of disks.}
\label{fig:Apollo}
\end{figure}

These ideas lead to a hierarchical picture of the energy landscape, where
large basins of attraction are surrounded by smaller basins, which in turn are 
surrounded by smaller basins, \ldots. 
A useful way to envisage this heterogeneity of basin areas is by analogy to an 
Apollonian packing, a two-dimensional example of which is shown in 
Fig.\ \ref{fig:Apollo}.
The initial configuration for this packing is three mutually touching discs. 
In the interstice between these three discs a new disc is added that itself 
touches the three discs (the so-called inner Soddy circle). In the three new 
interstices created, three more discs are added. This procedure is continued
iteratively to generate a space-filling packing of discs that is a 
fractal\cite{Mandelbrot} of dimension 1.3057.\cite{Manna91}

As with the basins of attraction that divide configuration space, the discs in
the Apollonian packing have a very heterogeneous distribution of areas 
that is organized hierarchically.  
In fact, the probability distribution of the areas of 
the disks is a power-law.\cite{Manna91} 
Furthermore, 
if we consider the Apollonian packing as a network where each 
disc corresponds to a node and nodes are connected by an edge if the 
corresponding discs are in contact, 
then this network has a scale-free degree 
distribution.\cite{Andrade04,Doye04e}
Moreover, many of the topological properties of this ``Apollonian network''
are similar to the inherent structure networks.\cite{Doye04e}

This analogy helps us to understand why our initial intuition---namely,
that the inherent structure networks would be lattice-like, because they are 
based on the adjacency of basins---was wrong. However,
there are of course limitations to the analogy. In the Apollonian packing
space is filled by an infinite number of discs, whereas configuration space
is divided up into a finite number of basins surrounding the minima.
Therefore, it's probably better to compare the ISN to an Apollonian packing 
where the iterative addition of discs has only been applied a finite number
of times.\cite{Doye04e} 

Having outlined the basic properties of the inherent structure networks, it
is useful to compare our results to the other studies looking at the 
connectivity of configuration space.
The scaling behaviour of $l_{\rm ave}$ has been characterized in both 
Scala {\it et al.}'s study of lattice polymers\cite{Scala01b} and 
Wuchty's analysis of RNA.\cite{Wuchty03}
In Ref.\ \onlinecite{Scala01b} networks of different size were generated by
considering both polymers of different length and with different
end-to-end distances. Based on a comparison of $l_{\rm ave}$ to that 
for a two-dimensional lattice and a random graph, they concluded that there is
a crossover to logarithmic scaling for networks of size larger than 100. 
However, this is the wrong comparison, because the dimension of configuration 
space is of course, $2N$, where $N$ is the number of monomers in the chain. 
Instead, the deviation from a power-law scaling is probably due to the 
effective higher dimensionality of conformation
space for the longer polymers. Consistent with this interpretation
the degree distribution shows no sign of heterogeneities, 
but is instead a Gaussian,\cite{Amaral00} because of the local nature of 
the move set.
Each conformation occupies a similar volume in conformation space.

Wuchty considers only one RNA sequence of a fixed length, 
but generates networks of different sizes through only including states 
that are within a certain variable energy of the ground-state. 
In contrast to our graphs, the networks are therefore incomplete.
$l_{\rm ave}$ deviates significantly from that for a random graph, and
the degree distribution has a exponential tail.

We would suggest that both these networks are essentially behaving as
lattices of high dimensionality, because the local move-sets only allow 
connections to conformations that are nearby in configuration space, and
there is no equivalent to the inherent structure mapping to group
conformations into states with widely-differing basin areas.

By contrast, in their study of a protein network Rao and Caflisch found
behaviour much more similar to our inherent structure networks with 
a degree distribution that had a power-law tail and with the low-lying
free-energy minima having greater connectivity.\cite{Rao04} This reflects
the conceptual similarities in the network representations noted in 
Section \ref{sect:eland_net} and
suggests the potential generality of the results we have found here.

\subsection{Randomization}

Before we look at more detailed properties of the networks, it is useful
to have an appropriate random model to which to compare. For example,
Dorogovtsev and Mendes have suggested that the values of $C/C^{\rm ER}$ 
for our inherent structure networks simply reflect the degree distribution 
and not any additional local structuring.\cite{Dorogovtsev03}

Analytic expressions are available for some network properties of a random 
ensemble of networks with a given degree distribution.\cite{Dorogovtsev03} 
However, these do
not apply when multiple edges and self-connections are excluded. Instead,
we used the switching algorithm to generate the ensembles of random 
networks.\cite{Maslov02,Milo04} 
At each step in this procedure two edges are picked at 
random, A---B and C---D say, and then rewired to form two new 
edges, either A---C and B---D, or A---D and B---C. 
It is evident that this rewiring conserves the degree distribution.
It is also straightforward to keep the constraint that there are no 
multiple edges or self-connections; rewiring steps that would break this 
constraint are simply not accepted. 
The procedure is started from the original network, and, after a suitable 
equilibration period, statistics were recorded. 
At each step the probability of each edge occurring is followed.  
From this probability distribution a variety of network properties can 
be calculated. However, for other properties, e.g.\ $l_{\rm ave}$,
that have to be explicitly calculated for a specific network, averages over 25 
independent (as judged by the autocorrelation function of the clustering 
coefficient) random networks were taken.

\begin{table}
\caption{\label{table:random}
Some properties of the randomized networks produced 
from the inherent structure networks by the switching 
algorithm.\cite{Maslov02,Milo04}}
\begin{ruledtabular}
\begin{tabular}{ccccc}
$N$ & $l_{\rm ave}$ & $C_1$ & $r_k$ & $r_E$ \\
\hline
 8 & $1.479$ & $0.625$ & -0.408 & -0.299 \\
 9 & $1.706$ & $0.492$ & -0.304 & -0.125 \\
10 & $2.043$ & $0.332$ & -0.158 & -0.088 \\
11 & $2.071$ & $0.219$ & -0.108 & -0.063 \\
12 & $2.247$ & $0.124$ & -0.115 & -0.085 \\
13 & $2.338$ & $0.071$ & -0.106 & -0.109 \\
14 & $2.302$ & $0.032$ & -0.068 & -0.119 \\
\end{tabular}
\end{ruledtabular}
\end{table}

Some properties of these randomized networks are given in Table 
\ref{table:random}.
For example, Figure \ref{fig:SW}(b) shows that although the randomized networks 
have a significantly higher clustering coefficient than the 
Erd\H{o}s-Renyi random graphs, the values for the inherent 
structure networks are yet still higher. This result disproves
Dorogovtsev and Mendes' assertion\cite{Dorogovtsev03} 
and shows that there are additional sources of local structuring, 
over and above that due to the degree distribution. 

By contrast, in Figure \ref{fig:SW}(a),
which compares $l_{\rm ave}$ for the cluster 
networks and their randomized versions,
only very small differences are apparent, showing that this property
primarily reflects the degree distribution of the networks. 
However, it is noticeable that $l_{\rm ave}^{\rm rand}$ is always
systematically smaller, albeit by a small amount, except for LJ$_{8}$.
This reflects the general correlation between higher clustering and
longer path lengths, as seen, for example, in the extremes of the
Erd\H{o}s-Renyi random graphs (low $C$, small $l_{\rm ave}$) and 
lattices (high $C$, large $l_{\rm ave}$). 
In a network with higher clustering, edges are `wasted' 
connecting nodes that are already close, which might have
been used to connect distant nodes and hence lower $l_{\rm ave}$
more significantly.  

In the following we will only show data for the randomized networks
when they show significant differences from the inherent structure networks. 
Typically, there will only be very small differences for properties related
to the shortest paths on the networks, as with $l_{\rm ave}$, 
but more significant differences appear for properties associated with 
clustering and network correlations.

\begin{figure}
\includegraphics[width=8.6cm]{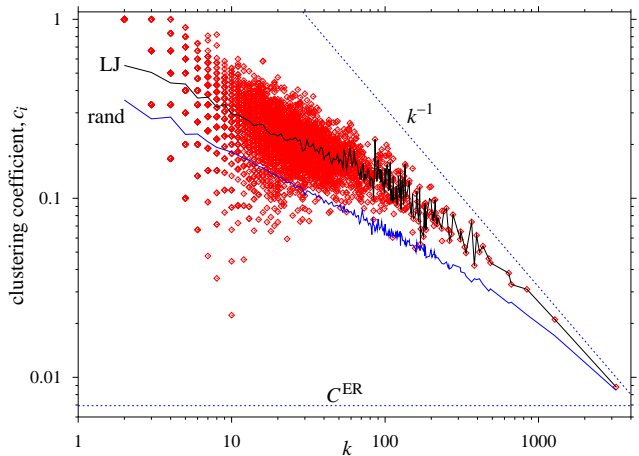}
\caption{(Colour online). 
The dependence of the local clustering coefficient of a node
on the degree. The data points are for each LJ$_{14}$ minimum, and the
the solid lines are the average values for a given $k$. For comparison,
the function $k^{-1}$ and  $c(k)$ for the randomized networks and 
an Erd\H{o}s-Renyi (ER) random graph have also been plotted.
 }
\label{fig:c_i}
\end{figure}

\subsection{Degree dependent properties for LJ$_{14}$}

The extremely heterogeneous nature of our networks is reflected in 
their scale-free degree distribution. 
To probe this heterogeneity further, 
it is also helpful to look at how properties vary from node to node, 
and particularly how they depend on the degree of a node. 
We will do this for our largest network, LJ$_{14}$, 
but results for smaller networks are similar.
Because of the large size of this network, there is often considerable 
scatter in the data. To make the trends more clear, averages of 
the properties, e.g.\ over nodes with the same degree, are also usually
presented. 

The dependence of the local clustering coefficient on degree is 
shown in Figure \ref{fig:c_i}. Like many other networks their is a clear 
association between higher degree and lower 
clustering.\cite{Vazquez02,Ravasz02,Ravasz03,Serrano03,Rao04,Cancho04} 
This property has often been attributed to a hierarchical 
structuring of the network.\cite{Ravasz02,Ravasz03} 
Notably many of the deterministic scale-free 
networks,\cite{Dorogovtsev02,Ravasz03,Comellas04}
including the Apollonian network mentioned in Section 
\ref{subsect:degree},\cite{Doye04e} follow 
the power law $c_i(k_i)\sim k^{-1}$.
At large $k$ the LJ$_{14}$ network roughly follows this law, 
but at low $k$ the $c_i$ values do not increase as fast as expected by
this law.

It is noteworthy that $c_i(k)$ for the randomized
networks has a similar type of the dependence on $k$, as 
for the LJ clusters. This is somewhat surprising since for a random
uncorrelated graph in which multiple edges and self-connections are allowed, 
the local clustering coefficient is independent of degree, no 
matter what the degree distribution.\cite{Dorogovtsev04}
The behaviour of $c^{\rm rand}_i(k)$ is due to the correlations induced by the 
absence of multiple edges and self-connections\cite{Maslov02b,Park03} 
and is something we will return to in the next section.

At small $k$ the local clustering coefficient for the ISN is roughly 50--100\%
larger than for the randomized networks. This extra local structuring
most likely reflects the spatial embedding of our networks.
Low degree nodes have small basins of attraction, and so are only 
connected to basins that are close in configuration space. This 
spatial localization of the connections leads to high clustering. 
By contrast, the large degree nodes have large basins of attraction, 
and so the minima surrounding them can be distant in configuration space,
and are not especially likely to also be connected to each other. 
Indeed, the clustering coefficient for the global minimum is similar to the 
value expected for an Erd\H{o}s-Renyi random graph.

\begin{figure}
\includegraphics[width=8.6cm]{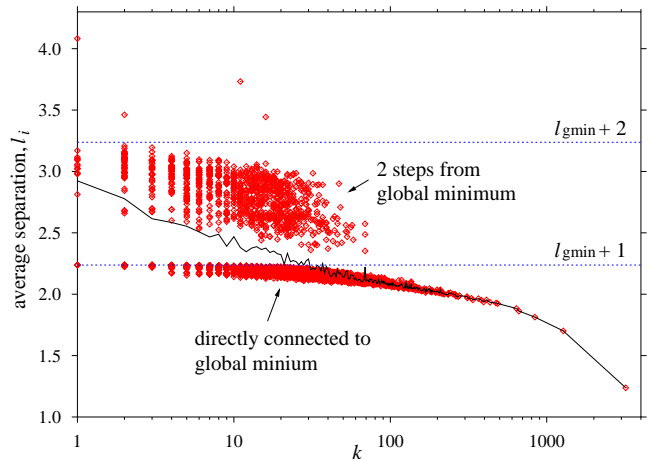}
\caption{(Colour online). 
The dependence of the average separation of a node in the network
from all the others on the degree.
Each LJ$_{14}$ minimum is represented by a data point and the solid line is 
the average value for a given $k$. $l_{\rm gmin}$ is the average separation
between the global minimum and the rest of the nodes.}
\label{fig:d_i}
\end{figure}

In Figure \ref{fig:d_i} we show $l_i$, the average separation of a 
node $i$ from the rest of the network. 
It is clear that the global minimum plays a
central role in the network. It is the node that is on average
closest to all the other nodes in the network. Similarly, 
for all our networks the global minimum
is one of the nodes for which all other nodes are within the
radius of the network (Table \ref{table:LJ}).
By contrast, low degree nodes are more on the periphery 
of the network. It is also apparent from Fig.\ \ref{fig:d_i} that
the minima separate themselves into sets depending on how far they
are away from the global minimum. For those that are directly connected to the
global minimum, $l_{\rm gmin}+1$ is an upper bound to $l_i$ and $l_i$ would
take that value if all the shortest paths involving that node passed through 
the global minimum. Similarly, the upper bound for those two steps away
from the global minimum is $l_{\rm gmin}+2$, and so on. In fact there are
only a few minima that are three steps away from the LJ$_{14}$ global minimum.

\begin{figure}
\includegraphics[width=8.6cm]{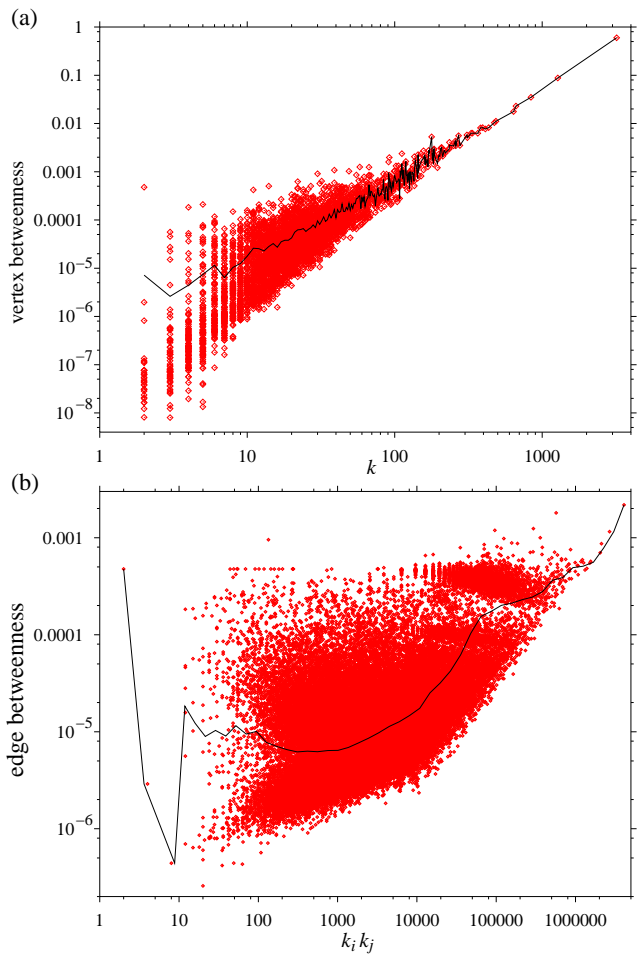}
\caption{(Colour online). 
The dependence of (a) the vertex betweenness on the degree
and (b) the edge betweenness on the product of degrees for LJ$_{14}$. Each
data point represents (a) a minimum or (b) an edge, and the solid line 
the average value for a given (a) $k$ or (b) range of $k_i k_j$.}
\label{fig:b_i}
\end{figure}

The vertex betweenness measures the fraction of the shortest
paths between all nodes in the network that pass through 
a particular vertex. Similarly the edge betweenness measures
the fraction of these paths that pass along a particular 
edge in the network. These quantities provide a measure 
of the importance of a node or edge to a network, particularly 
for dynamical processes that occur on the network, 
such as the passage of information.
For scale-free networks, there is typically a strong correlation 
between the vertex betweenness and the degree,\cite{Holme02}
and Figure \ref{fig:b_i}(a) shows that the inherent structure
networks show a similar behaviour.
The vertex betweenness has roughly a power-law dependence on $k$, and
so given the scale-free degree distribution, this also implies
that the probability distribution for the betweenness also has 
a power-law tail.\cite{Goh02}
Again the global minimum plays a central role in the network. 
For LJ$_{14}$ 59.8\% of the shortest paths pass through this node.

For scale-free networks, there is also usually a correlation 
between the edge-betweenness and some measure of the degrees at
either end of the edge, but it is much weaker than for the vertex 
betweenness.\cite{Holme02} 
In Figure \ref{fig:b_i}(b) we see such a correlation with the product
of the degrees at either end of the edges for the LJ$_{14}$ network.
However, the correlation is fairly weak and there is much more 
scatter in the data. The edge with maximum betweenness connects the 
two highest-degree nodes and 0.22\% of the shortest paths pass along it.

\subsection{Network Correlations}

To further understand the structure of our networks, it is important 
to go beyond just local properties of a node to study the correlations between
the properties of the nodes. Indeed correlations can have an important 
influence on other network properties,\cite{Dorogovtsev04} and have 
a strong effect on dynamical processes that occur on a 
network.\cite{Eguiluz02,Boguna03b,Vazquez03,Vazquez03b,Echenique04}

\begin{figure}
\includegraphics[width=8.6cm]{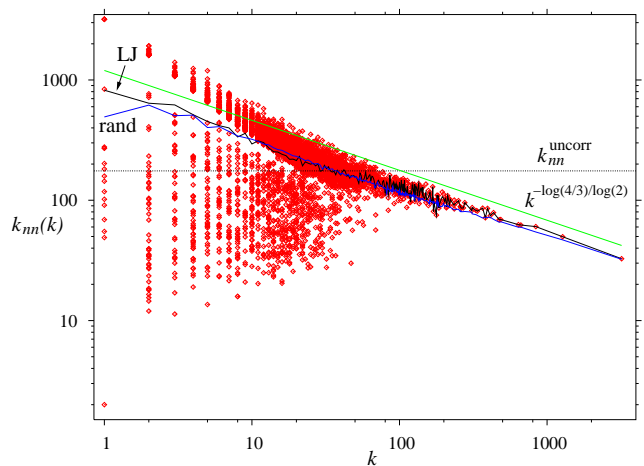}
\caption{(Colour online). 
The dependence of the average degree of the neighbours of a node 
on the degree of that node. Each LJ$_{14}$ minimum is represented by a 
data point and the solid line labelled `LJ' is the average value 
for a given $k$.
Also included is $k_{nn}(k)$ for the randomized networks, a line with 
the exponent expected for an Apollonian network,\cite{Doye04e} and a line at 
$k_{nn}^{\rm uncorr}$, the value expected for an uncorrelated network.}
\label{fig:knn}
\end{figure}

Most commonly, correlations with respect to degree are investigated. 
For example, in Figure \ref{fig:knn} we show how $k_{nn}$, the average 
degree of the neighbours a node, depends on the degree of the node.
If the network were uncorrelated, one would expect $k_{nn}$ to be independent
of degree and to take a value 
$k_{nn}^{\rm uncorr}=\langle k^2\rangle/\langle k\rangle$.
However, $k_{nn}(k)$ has a negative slope showing that our networks
are disassortative by degree, i.e.\ high degrees are more likely to be
connected to low degree nodes than expected, and {\it vice versa}.
Conversely, in assortative networks nodes are more likely to be connected
to those with similar degrees.
Interestingly, $k_{nn}(k)$ has an approximate power-law dependence on $k$,
as seen for other scale-free networks,\cite{Pastor01b} with an
exponent that is very similar to that of an Apollonian network.\cite{Doye04e}

Another way to measure such correlations is through the assortativity 
coefficient introduced by Newman.\cite{Newman02a,Newman03b}  
It is a two-point correlation function of the 
properties at either end of an edge and is usually normalized by 
the value expected for a perfectly assortative network.
It is defined as
\begin{equation}
r_s={\langle s t\rangle_e - \langle s\rangle_e\langle t\rangle_e \over
    \langle s t\rangle_{e,assort} - \langle s\rangle_e \langle t\rangle_e }.
\label{eq:assort}
\end{equation}
where $s$ and $t$ correspond to the property of interest at
either end of an edge, $e$ denotes that the averages are over all edges
and {\it assort} that the average is for a perfectly assortative network.
A positive value is expected for an assortative network (with an upper bound
of 1), zero for an uncorrelated network and a negative value for a 
disassortative network. Negative values have been found for technological
networks, such as the internet and world-wide web, and biochemical networks,
such as the network of protein-protein interactions and metabolic networks,
whereas strongly positive values have been found for social 
networks.\cite{Newman02a,Newman03b}  
Values of $r_k$ are given in Table \ref{table:LJ} and, as with
$k_{nn}(k)$ indicate the disassortative nature of our networks.
The values found for the larger clusters are quite similar to those
for technological networks.\cite{Newman02a,Newman03b}  

The origin of this disassortativity
has been explored in most depth for the 
internet.\cite{Pastor01b,Maslov02b,Park03} 
In particular, it was found that the exclusion of 
multiple edges and self-connections can be a significant source of 
disassortativity. Indeed, when we compare $k_{nn}$ and $r_k$ to the values
for the randomized networks (Fig.\ \ref{fig:knn} and Table 
\ref{table:random}), we find very similar levels of disassortativity.

In an uncorrelated network, the probability that a randomly chosen edge will
connect two particular nodes is $k_1 k_2/2 n_{\rm edges}^2$ 
and, hence, the total
number of expected connections between the 
two nodes is $k_1 k_2/2 n_{\rm edges}$.
For the LJ$_{14}$ network this gives an expected value of 33.5 edges between
the two highest degree nodes.
As only one is possible, the network appears disassortative.

Correlations also have a significant effect on the 
clustering.\cite{Dorogovtsev04,Soffer04}
As indicated by the expressions above, the probability that the neighbours 
of a node are connected depends sensitively on the degrees of the neighbours,
with high degree nodes being  much more likely to be connected. 
Hence the local clustering coefficient is
higher for nodes with larger $k_{nn}$, and so even for the
randomized graphs the local clustering coefficient decreases as $k$ 
increases (Fig.\ \ref{fig:c_i}).

\begin{figure}
\includegraphics[width=8.6cm]{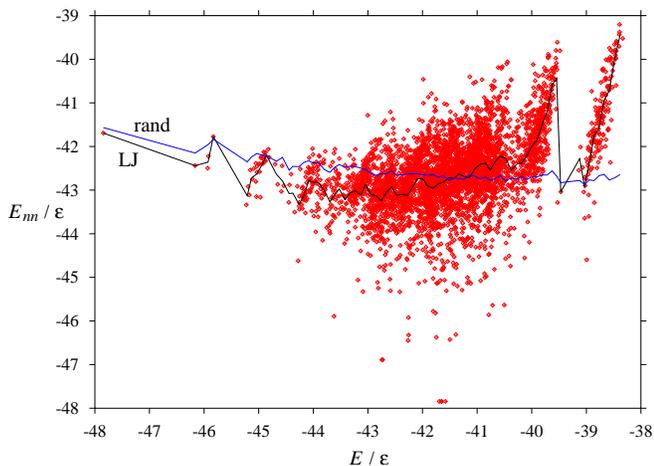}
\caption{(Colour online). 
The dependence of the potential energy of the neighbours
of a minima on its energy. The data points are for every LJ$_{14}$ 
minimum, and the lines binned averages for both the LJ$_{14}$ 
network, and randomized versions, as labelled.}
\label{fig:Enn}
\end{figure}

We have also explored correlations between the potential
energies of connected minima. Values of $r_E$ are given for both the 
inherent structure networks
and the randomized versions in Tables \ref{table:LJ} and \ref{table:random}, 
respectively, and the behaviour of $E_{nn}$, 
the average energy of the the minima
to which a minimum is directly connected, is shown in
Figure \ref{fig:Enn}. Because of the correlations
between $k$ and $E$ (Fig.\ \ref{fig:LJ14_kE}), 
one might also expect the networks to 
be disassortative with respect to the energy of the minima.
However, for $N>9$ positive values of $r_E$ are found, whereas the values
for the randomized networks are always negative, and of roughly similar 
magnitude to $r_k^{\rm rand}$.
This difference shows that the potential energy of the minima plays an
important role in the structure of the network. 

At low energies, $E_{nn}(E)$ has a gentle negative slope similar to 
that for the randomized graphs, and probably mainly reflects the 
exclusion of self-connections and multiple edges. The low-enegy hubs
have to be connected to higher-energy minima because of the constraint
that all the neighbours must be different.
For example, the LJ$_{14}$ global minimum is connected 
to 3201 of the 4196 nodes. 
Even if these were the 3201 lowest-energy nodes (i.e.\ maximally assortative
given the exclusion of multiple edges),
the average energy of its neighbours would only be a little 
lower ($0.28\epsilon$) than the actual value.
However, for the higher-energy nodes $E_{nn}(E)$ has 
a strongly positive slope, because of the preference for linking to
minima with similar energy.

Although uncorrelated energy landscapes have been 
studied theoretically,\cite{Derrida80,Bryngelson87} 
the assumption that the energy of connected states 
is independent is usually unrealistic and gives rise to landscapes
that are extremely rough.
(Indeed the first step in incorporating greater realism in these models
is often to include correlations.\cite{Saven94,Wang96})
For landscapes associated with the configurations of atomic
systems, the landscapes are much smoother, simply because connected
states are likely to be structurally similar and hence 
have similar energies. 
However, it should be remembered that the measures of correlations we have used
here are quite local, and apply to just the immediate neighbourhood of a basin. 
When energy landscapes are
inferred to be rough, e.g.\ because the system is a good glass-former and
easily gets stuck in traps on the landscape,\cite{Still95} this is 
usually referring to a lack of correlations at larger length scales.

\section{Dynamical Implications}
\label{sect:discuss}

So far, we have made a detailed characterization of the network of connections
between the minima on a potential energy surface, and
these results provide important insights into the fundamental organization
of energy landscapes. But it is also important to understand the implications
of this topology for the dynamics of a system.

Clearly, the small-world nature of the inherent structure networks
has important consequences for the theoretical understanding of searching 
configuration space.
For example, the Levinthal paradox in protein folding highlights the 
huge number of possible configurations of a protein and the seeming 
impossibility of reaching the native state in a random search through 
these configurations, in contrast to the protein's actual ability to 
fold.\cite{Levinthal_Mark}
Similar paradoxes can of course be formulated for any system with 
a large configuration space, be it a macromolecule, cluster or bulk liquid,
that is able to routinely find a preferred structure.
More formally, the exponential increase in the size of the search space 
underlies the classification of most problems involving the
global optimization of a configuration as ``NP-hard'';\cite{Garey} 
i.e.\ there is no known algorithm that is guaranteed to solve the problem in
polynomial time.

Our results offer part of the answer to these paradoxes. Even for
high-dimensional configuration spaces with huge numbers of 
accessible states, the number of steps in a pathway from a 
random configuration to the target state remains modest
due to the favourable scaling with size (Eq.\ (\ref{eq:lave_land})).
Of course, finding that particular pathway in the absence 
of additional guiding information 
can be extremely hard, although there is some evidence that a local
knowledge of the topology can be of some help.\cite{Adamic01,Kim02} 
Indeed, the emphasis of most proposed solutions to the
Levinthal paradox is on those features of the energy landscape,
such as funnels,\cite{Leopold,Bryngelson95} that guide the system 
in the right direction in its descent down the energy landscape, 
and whose presence differentiates those systems that are able to 
find the target state,
from those that get stuck in the morass of possible configurations. 

The dramatic consequences of this small-world behaviour are
evidenced in examples where the energy landscape has a favourable topography.
For example, for LJ$_{55}$ it has been estimated that there are $10^{21}$ 
minima. However, the basin-hopping global optimization algorithm is able to
find the global minimum from a random starting point after sampling on average
only approximately 150 minima.\cite{Doye98e}

The scale-free nature of the inherent structure network will also have 
important implications for a system's dynamics.
Interestingly, at the centre of Leopold {\it et al}'s original definition 
of a folding funnel
was the concept of a ``convergence of multiple pathways''
at the native state of a protein.\cite{Leopold} Interpreting
this idea in network terms, it seems to be suggesting  that the native state 
would be a hub in the network. For our inherent structure networks we see 
such convergence with the global minimum connected to a
significant fraction of the other minima, 
and similar results have been obtained in 
Rao {\it et al.}'s investigation of the connectivity of configuration space
for polypeptide chains.\cite{Rao04} Of course, there is also a strong 
topographical component inherent in the idea of a funnel, 
namely that as the system goes downhill it is becoming closer to the 
native state, and it is this feature that has received the most 
emphasis more recently, perhaps to the neglect of the potential 
focussing aspects of the topology.

A wide variety of dynamical processes have been studied 
on complex networks. The dynamics of interest here is the molecular dynamics
on the potential energy surface, as dictated by the forces on the system.
The most relevant work to this from the networks literature is 
that examining diffusion on scale-free networks,\cite{Bollt04}
where it was found that this topology led to more favourable scaling
of the transit time between nodes, particularly when these nodes 
had high-degree.

The effect of the topology on the dynamics on an energy landscape
can be illustrated for a simple model landscape where topographical 
features are eliminated. 
The model consists of a a flat landscape where all the minima are identical, 
except for their connectivity, 
i.e.\ they have the same energies and vibrational 
(and hence thermodynamical) properties, and 
furthermore, all the transition rates between
connected minima take the same value $k^\dagger$.
The average residence time in a minimum $i$ is then 
$1/k_i k^{\dagger}$. As the equilibrium probability of 
being in a minimum $i$ is simply $1/n_{\rm min}$, the
frequency with which this minimum is visited is $k_i k^\dagger/n_{\rm min}$. 
The first passage time for encountering a minimum therefore decreases 
with degree. The topology of the inherent structure network has
a focussing effect directing the system more rapidly to 
the highly-connected hubs. The effects of the topology on the dynamics
will be explored in more detail in future work.

\section{Conclusion}

In this paper we have provided 
a detailed characterization of the inherent structure networks
for a series of small Lennard-Jones clusters. 
These networks show the mixture of local order (as measured by 
clustering coefficients) and small average separations between nodes that
is characteristic of small-world networks.\cite{Watts98} 
Furthermore as the size of the clusters increase, 
these inherent structure networks develop a clear power-law tail to the
degree distribution and have a whole variety of properties that are typical
of scale-free networks.

However, in contrast to most scale-free networks the origin of this 
scale-free topology is not due to some form of preferential attachment 
during network growth.\cite{Barabasi99}  
Instead, the network heterogeneity reflects the topography of the energy 
landscape with the degree of a node strongly correlated to the potential 
energy of the associated minimum. This correlation most likely arises
because the low-energy minima have larger basins of attraction, and
so are likely to have more transition states located on their basin boundaries.
Apollonian networks provide a model of how such basins could be organized
to give rise to a scale-free network.
Hence, the discovery of the scale-free character of inherent structure networks
may have profound implications for our understanding of the way 
the inherent structure mapping divides up configuration space, 
and raises the possibility that configuration space is tiled 
by a fractal, Apollonian-like packing of basins.

The results presented here for the inherent structure networks 
are for systems interacting with a particular potential and of small size, 
so it is natural to ask how general are the results.
Firstly, we see no obvious reason why the Lennard-Jones potential should be 
``special'', producing scale-free networks, while the landscapes
associated with other potentials have different topologies.
However, it is conceivable that differences might arise as a result of 
the orientational degrees of freedom associated with molecular 
(not atomic) systems or the constraints of 
chain connectivity associated with polymers, and we intend to explore this
further in future work.
The work of Rao and Caflisch is also of interest in this regard, because the 
networks they constructed to represent the configuration space connectivity 
of polypeptides also showed scale-free behaviour.\cite{Rao04}

Secondly, because of the exponential increase in the number of minima 
and transition states with system size,
the analysis we presented here is inevitably limited to relatively small sizes.
Furthermore, there is currently no known method to construct a statistical
representation of the inherent structure network from an incomplete 
sampling of the network.
One potential way to analyse systems of larger size is to use a 
more coarse-grained division of configuration space than 
provided by the inherent structure mapping, such as that used by 
Rao and Caflisch.\cite{Rao04} 
Another more indirect approach would be to probe properties that 
sensitively reflect the underlying network topology. For example, 
the analogy to Apollonian networks suggests that the distribution
of basin areas should follow a power law with an exponent
approximately equal to -2.\cite{Doye04e}


\begin{thebibliography}{93}
\bibitem[{Tx()}]{Tx}
\bibinfo{note}{The temperature at which this time scale separation disappears
  was labelled $T_x$ by Goldstein\cite{Goldstein69} and has been found to be
  near to the mode-coupling temperature.\cite{Schroder00}}
\bibitem[{PRL()}]{PRLbug}
\bibinfo{note}{In Ref.\ \onlinecite{Doye02c} we gave the result $l_{\rm
  ave}^{\rm latt}=N(L+1)$. The error in our calculation was to exclude terms
  with $i=j$ in Eq.\ \ref{eq:l1D}, before multiplying by $3N$. However, the
  result of this is to exclude from the average all paths between nodes that
  have the same value for one of the lattice coordinates, rather than just
  between a node and itself, as we had intended. This mistake did not affect
  the main point of the calculation, the logarithmic scaling of $l_{\rm
  ave}^{\rm latt}$ with the number of lattice points, but just the prefactor to
  Eq.\ \ref{eq:lave_latt}.}
\bibitem[{Tsa()}]{Tsallisfit}
\bibinfo{note}{As noted by Tsallis and co-workers\cite{Tsallis04,Soares04} the
  cumulative degree distribution can in fact be well fitted to a function of
  the form $e_q^{-\beta k^2}$, where the $q$ exponential is defined as
  $e^x_q=[1+(1-q)x]^{1/(1-q)}$; note that $e_1^x=e^x$.}
\end{thebibliography}
\end{document}